\documentclass{article}
\usepackage{spconf,amsmath,graphicx}
\usepackage{bm}
\usepackage{longtable, array, booktabs}
\usepackage[table]{xcolor}
\usepackage[colorlinks, linkcolor=black,
    filecolor=black,      
    urlcolor=black,
    citecolor=black]{hyperref}

\title{Towards high-fidelity singing voice conversion with acoustic reference and contrastive predictive coding}

\name{
Chao Wang$^1$$^2$, Zhonghao Li$^2$, Benlai Tang$^2$, Xiang Yin$^2$, Yuan Wan$^2$, Yibiao Yu$^{1}$,
Zejun Ma$^2$,
}
\address{
\normalsize  $^1$Soochow University \\
\normalsize  $^2$ByteDance AI Lab \\
\normalsize georgefly1@163.com, \{lizhonghao.01,tangbenlai\}@bytedance.com
}
\begin{document}

\fontsize{9.35pt}{\baselineskip}\selectfont 

\maketitle

\begin{abstract}
Recently, phonetic posteriorgrams (PPGs) based methods have been quite popular in non-parallel singing voice conversion systems. However, due to the lack of acoustic information in PPGs,  style and naturalness of the converted singing voices are still limited. To
solve these problems, in this paper, we utilize an acoustic reference encoder to implicitly model singing characteristics. We experiment with different auxiliary features, including mel spectrograms, HuBERT, and the middle hidden feature (PPG-Mid) of pretrained automatic speech recognition (ASR) model, as the input of the reference encoder, and finally find the HuBERT feature is the best choice. In addition, we use contrastive predictive coding (CPC) module to further smooth the voices by predicting future observations in latent space. Experiments show that, compared with the baseline models, our proposed model can significantly improve the naturalness of converted singing voices and the similarity with the target singer. Moreover, our proposed model can also make the speakers with just speech data sing. 
\end{abstract}


\section{Introduction}

\label{sec:intro}
Singing voice conversion is intended to change the timbre of singing voices while keeping the musical contents like melody and lyrics maintained. It has achieved more and more attention in recent years, since it can be used in many aspects, like improving the singer's vocal quality and creating a virtual singer. 

The early studies are mainly focused on parallel singing voice conversion, which means there are parallel training data between the source singer and the target singer. These traditional methods\cite{ref1}\cite{ref2}\cite{ref3} use statistical models to build the mapping between the parallel samples. However, it's quite expensive to collect parallel samples of different singers in real application scenarios. So recently, the studies of non-parallel voice conversion have been the mainstream. The key idea of non-parallel singing voice conversion is to extract the singer-independent content representation first, and then generate the converted singing voices using the target singer embedding. According to whether the singer-independent content representation is extracted in an unsupervised manner, non-parallel singing voice conversion can be categorized into auto-encoder based and PPGs based methods.

Auto-encoder based methods disentangle the content feature from speech in an unsupervised manner by imposing constraints on encoder output. Domain confusion \cite{ref4} module is used in \cite{ref5}\cite{ref6} to remove speaker-dependent information like timbre from speech. In \cite{ref7}\cite{ref8}, feature disentanglement is realized using variational auto-encoder (VAE). \cite{ref9} utilizes vector quantized variational auto-encoder (VQ-VAE) as an information bottleneck to get the content information. Although auto-encoder based methods can achieve acceptable results in non-parallel singing voice conversion, its ability of feature disentanglement is not robust enough for various singing input.

In PPGs based methods\cite{ref10}\cite{ref11} \cite{ref12}\cite{ref13}\cite{ref14}, they use the last layer output of the pretrained ASR model, namely PPGs, to represent content information. Since the ASR model is pretrained with huge amount of training data and the training targets are phoneme labels, its ability of eliminating acoustic information is quite robust. However, compared with speech, singing voice contains much more characteristics like prosody and rhythm, which are difficult to model explicitly only using PPGs. Several works have tried to supplement acoustic features in singing voice conversion. In \cite{ref13}, they use pitch to represent melody. However, pitch itself is still not enough to model singing characteristics. In \cite{ref14}, apart from pitch, they feed  mel spectrograms into reference encoder to model singing characteristics implicitly. However, as mentioned in auto-encoder based methods, there exists too much singer-dependent information in mel spectrograms, which is difficult to remove and will finally degrade the performance. 

\begin{figure*}[h]
\centering
\includegraphics[width=15cm,height=4.6cm]{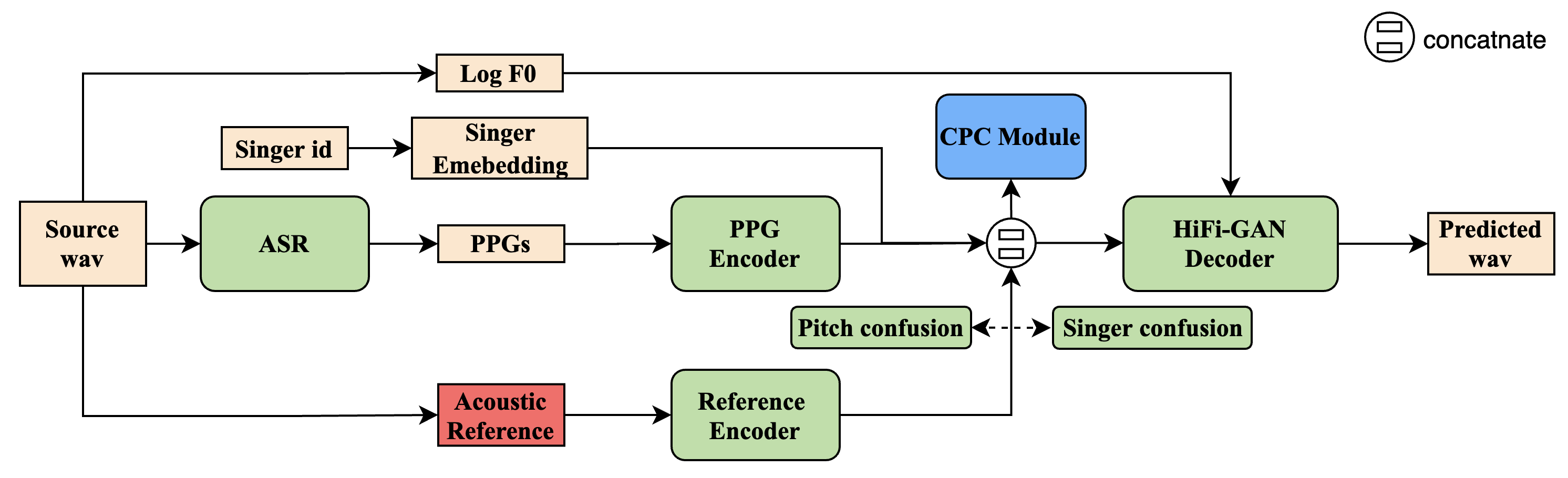}
\caption{Overall architecture of our proprosed method. Singer confusion, pitch confusion, and CPC module are not used in the inference phase.}
\end{figure*}

In this work, our contribution is two-fold: (1) We try to find an appropriate auxiliary feature that contains enough singer-independent acoustic and musical information to serve as the reference encoder input. Apart from mel spectrograms, We select HuBERT\cite{ref15}, which is a kind of feature originally proposed in self-supervised speech representation learning, and middle hidden feature of ASR model, denoted as PPG-Mid. Both HuBERT feature and PPG-Mid feature contain linguistic and acoustic information. Meanwhile, they contain less singer information than mel spectrograms. In the experiments, we compare these three features when they serve as the reference encoder inputs. To further eliminate the remaining singer-dependent information like timbre in the reference encoder output, both singer confusion and pitch confusion module are also used.  (2) To alleviate local errors and enhance the smoothness of voices, inspired by recent studies proposed in speech self-supervised representation learning\cite{ref16}\cite{ref17}\cite{ref18}, we utilize contrastive predictive coding (CPC) module which predicts future representations in the encoder output feature space. In \cite{ref19}, they use the CPC module to eliminate the local errors in the vector quantization output. In our paper, we are the first one to utilize the CPC module in the PPGs based singing voice conversion system. 

The rest of the paper is organized as follows:
Section 2 will present the proposed method. In Section 3, Experiments results are presented. Section 4 concludes this paper.
\vspace{-0.3cm}
\section{Methods}
\label{sec:majhead}
The overall architecture of our proposed method is shown in Fig.1. It is mainly composed of pretrained ASR model, PPG encoder, acoustic reference encoder, CPC module, and HiFi-GAN\cite{ref20} decoder.

The PPG encoder is to acquire singing content from PPGs feature. Meanwhile, the auxiliary acoustic feature of source voices is fed into the reference encoder. The outputs of PPG encoder and reference encoder are concatenated to form the final encoder output. Then encoder output, singer embedding, and pitch are fed into the HiFi-GAN decoder to reconstruct the speech waveform, where pitch is fused into the decoder in the same way as \cite{ref12}. Meanwhile, the singer and pitch confusion module are used to remove the singer information in the reference encoder output. The ASR model is based on Deep-FSMN\cite{ref21}, which is a heirarchical network with 38 layers. The model architectures of the PPG encoder and acoustic reference encoder are CBHG network proposed in \cite{ref22}. We use the HiFi-GAN vocoder proposed in \cite{ref20} as the decoder to reconstruct waveform directly. The reconstruction loss used for training the network follows \cite{ref20}.

In the following part, we will introduce the acoustic reference features, and the CPC module in detail.

\subsection{Acoustic Reference}
\label{ssec:subhead}
Singing voice contains rich acoustic and musical information. However, PPGs do not contain enough acoustic and musical information, which makes it difficult to generate high-fidelity singing voices. To model the acoustic and musical information for singing voice conversion, we incorporate a reference encoder. As for the reference encoder input, we select mel spectrograms, PPG-Mid, and HuBERT feature as the candidates: 

\textbf{Mel spectrograms}:
Although mel spectrograms contains rich acoustic information, too much singer-dependent information like timbre also exists in mel spectrograms, which will finally degrade the quality of the converted singing voices and reduce the similarity with the target speaker.

\textbf{PPG-Mid}:
PPG-Mid is the output feature of 30th hidden layer in DFSMN\cite{ref21} based ASR model. On one hand, Compared with PPGs, the middle hidden layer output PPG-Mid contains more acoustic information because it is the shallower layer. On the other hand, The singer information in PPG-Mid is much less than that in mel spectrograms, because ASR model has already eliminated most of the singer information by linguistic label training.

\textbf{HuBERT}:
HuBERT feature\cite{ref15} is a kind of self-supervised speech representation, which achieves the state-of-the-art results in the ASR task. As shown in Fig.2, it utilizes an offline clustering step to provide aligned pseudo target labels for prediction. Because audio data of different speakers are trained towards the same pseudo target labels, most of the speaker information in the HuBERT feature are removed. The prediction loss is over the masked regions only, forcing the model to extract high-level representations from the unmasked frames to infer the masked targets correctly. Therefore, the HuBERT feature contains a combined acoustic and linguistic information.
 
However, the above three acoustic reference features still keep some singer information like pitch and timbre, so we still need the singer confusion and pitch confusion module like in \cite{ref6} to eliminate these information. The reference encoder output is fed into the singer-identity classifier and pitch predictor. Both the confusion modules consist of four 1-D convolution layers activated by ReLU layer followed by one final linear projection layer. The singer classifier is trained using cross entropy loss $L_{s}$ to correctly classify the speaker label, and pitch predictor is optimized using MSE loss $L_{f}$ to predict the pitch. The confusion loss $L_{confusion}$ to update the reference encoder is defined as below: 
\begin{equation}
        L_{confusion} = -\lambda*L_{s} - \omega*L_{f} 
\end{equation}

where scalar $\lambda$ and $\omega$ are the weight of the singer confusion loss and pitch confusion loss, respectively.

\begin{figure}[h]
\centering
\includegraphics[width=8cm,height=3.9cm]{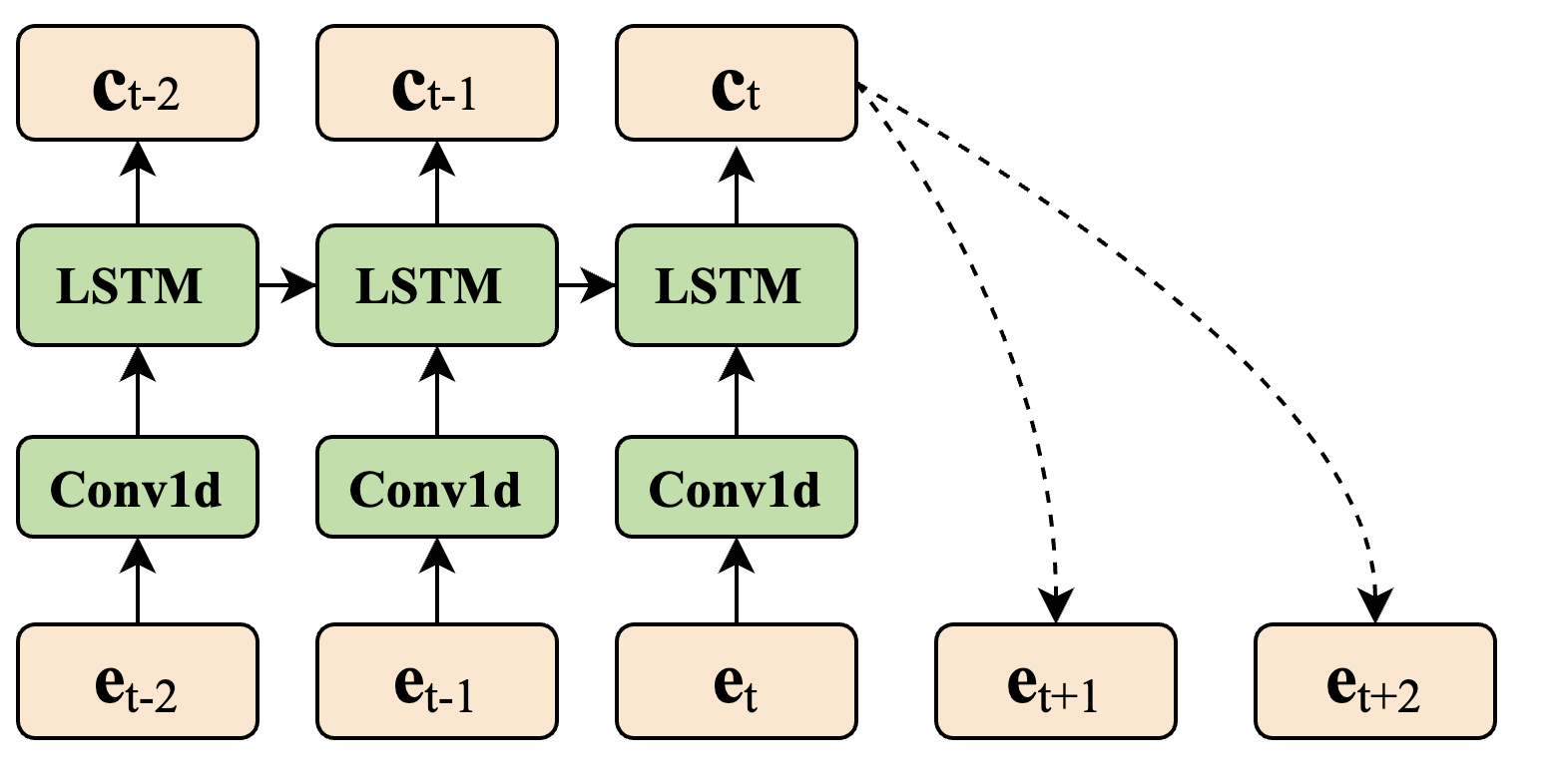}
\caption{The architecture of the CPC module}
\end{figure}

\vspace{-0.38cm}
\subsection{CPC module}

The quality of the converted singing voices depends on the accuracy of the pretrained ASR model. The local recognition errors in the PPGs feature will bring noise in the converted singing voices and degrade the quality of singing voice conversion. In addition, although the singer and pitch confusion module can get rid of the remaining singer information like timbre in the reference encoder output, it inevitably harms the phonetic and vocal information at the same time, which will cause local instability in the converted singing voices.

To overcome these problems, we employ contrastive predictive coding (CPC) module as the regularization loss to predict  future representations  in  the encoder output  feature  space. The CPC module is originally proposed in the self-supervised learning\cite{ref16}\cite{ref17}\cite{ref18} to learn meaningful speech represention, which can be used in various downstream tasks. In this paper, we utilize CPC module to discard local noise and errors in the encoder output and enhance the voice smoothness. 

As shown in Fig.2, the encoder output $\bm{E} = \{\bm{e}_{1}, ... \bm{e}_{T}\}$ of length T is fed to the context network, which in our paper is composed of long short-term memory network (LSTM) and 1-D convolution layer, to get the context output $\bm{C} = \{\bm{c}_{1}, ... \bm{c}_{T}\}$. Given current timestep context feature $\bm{c}_{t}$, the CPC module needs to distinguish the postive future k-step feature $\bm{e}_{t+k}$ from the remaining negetive feature samples. If $\bm{e}_{t}$ contains vocal errors, then $\bm{c}_{t}$ also contains errors. Given $\bm{e}_{t+k}$ contains precise vocal information, the network learns to discard the local errors and noise in $\bm{e}_{t}$ and $\bm{c}_{t}$ so that $\bm{e}_{t+k}$ can be predicted correctly. The CPC loss term is as follow:
\begin{equation}
    L_{CPC} = -\beta*\sum_{t=0}^{T-t-1}\sum_{k=1}^{K} log[\frac{exp(\bm{e}_{t+k}^{T}\bm{W}_{k}\bm{c}_{t}))}{\sum_{\tilde{e}\in \Omega }^{}exp(\bm{\tilde{e}}^{T}\bm{W}_{k}\bm{c}_{t})}]
\end{equation}
    where T is the length of the encoder output, K is the largest timestep we predict, and $\beta$ is the CPC loss weight. $\bm{W}_{k} (k = 1, 2, ..., K)$ is a trainable projection matrix and $\Omega$ is the set containing the negative samples. In our paper, the negetive samples set is formed by randomly selected from the same utterance as the positive sample. The K is set to 12.

Let the  HiFi-GAN\cite{ref20} reconstruction loss be the $L_{hifi}$, which consists of Mel spectrograms loss, deep feature mathcing loss, and adversial loss. The overall generation loss of our proposed model is shown as below:
\begin{equation}
    L_{G} = L_{hifi} + L_{confusion}+ L_{CPC}
\end{equation}

\section{Experiments}
\label{sec:experiments}
\subsection{Experiment setup}
\label{ssec:subhead}
We evaluate our proposed model on a mixed training dataset which contains VCTK speech corpus\cite{ref23} and NUS-48E singing corpus\cite{ref24}. 90\% of the audios in datasets are selected for training, 5\% for validation. The test set
consists of 20 segments from 10 singers. For evaluation, we select a female singer and a male singer as the target timbres. Except for section 3.2.4, the test source audios and target singers for evaluation are selected from NUS-48E corpus, namely in-domain singing voice conversion. All songs are sampled at 24kHz. 80-dimensional mel spectrograms are extracted with a Hanning window of 40 ms and 10 ms frame shift.

 We use 20k singing data to pretrain the ASR model. The HuBERT features are extracted using the pretrained model provided in \cite{ref15}. We train our proposed model using Adam \cite{ref25} optimizer for 400K steps. The learning rate is 0.0004 and the batch size is 16. Weighting factors $\lambda$, $\omega$, and $\beta$ are all set to 0.1. We use pyreaper\footnote{\href{https://github.com/r9y9/pyreaper}{https://github.com/r9y9/pyreaper}} to extract the pitch from the audios. Samples can be found online
\footnote{\href{https://georgehappy1.github.io/svcdemo/}{https://georgehappy1.github.io/svcdemo/}}.

\subsection{Evaluation}
\label{ssec:subhead}
We invite 15 music professionals to evaluate the results. For subjective evaluation, we select 1-5 mean opinion score (MOS) (1-bad, 2-poor, 3-fair, 4-good, 5-excellent) for both singing naturalness and timbre-style similarity with the target singer. For objective evaluation, we select normalized cross-correlation (NCC) to measure the pitch matching degree between the source and converted singing audios. The higher the NCC score is, the more accurate the pitch is. 

\begin{table}[ht]
\caption{The evaluation results of different acoustic reference.}
\begin{tabular}{l c c c }

\specialrule{0.1em}{2pt}{2pt}
\multicolumn{1}{c}{\textbf{Method}}  & \textbf{Naturalness} & \textbf{Simiarity} & \textbf{NCC} \\ \specialrule{0.1em}{2pt}{2pt}
Ground\_truth    &    $4.76\pm{0.07} $  &   -- &  --  \\ \hline
Prop\_mel    &    $3.56\pm{0.16}$  &  $3.48\pm{0.16}$ & $0.976$  \\ \hline
Prop\_ppg-mid &    $3.49\pm{0.18}$  &  $3.58\pm{0.18}$ & $0.982$ \\ \hline
Prop\_hub &    $3.62\pm{0.17}$  &  $3.63\pm{0.16}$ &  $0.985$ \\ \specialrule{0.1em}{2pt}{2pt}
\end{tabular}
\label{tab:table1}

\end{table}

\begin{table}[ht]
\caption{The evaluation results of the ablation study for the CPC module}
\begin{tabular}{ l c c c }
\specialrule{0.1em}{2pt}{2pt}
\multicolumn{1}{c}{\textbf{Method}}   & \textbf{Naturalness} & \textbf{Simiarity} & \textbf{NCC} \\ \specialrule{0.1em}{2pt}{2pt}
Prop\_hub      & $3.62\pm{0.17}$  & $3.63\pm{0.16}$     &  $0.985$   \\ \hline
Prop\_hub\_cpc & $3.74\pm{0.19}$  &  $3.87\pm{0.17}$     & $0.987$    \\ \specialrule{0.1em}{2pt}{2pt}
\end{tabular}
\label{tab:table2}
\end{table}

\begin{table}[ht]
\caption{The evaluation results of baseline models and our proposed model }

\begin{tabular}{ l c c c }
\specialrule{0.1em}{2pt}{2pt}
\multicolumn{1}{c}{\textbf{Method}}   & \textbf{Naturalness} & \textbf{Simiarity} & \textbf{NCC} \\ \specialrule{0.1em}{2pt}{2pt}
Baseline1             & $2.49\pm{0.19}$  & $2.66\pm{0.19}$ &  $0.969$   \\ \hline
Baseline2             & $1.85\pm{0.13}$  &  $1.94\pm{0.13}$  &  $0.949$        \\ \hline
Prop\_hub\_cpc & $3.74\pm{0.19}$  &   $3.87\pm{0.17}$ &   $0.987$     \\ \specialrule{0.1em}{2pt}{2pt}
\end{tabular}

\label{tab:table3}
\end{table}

\begin{table}[ht]
\caption{The evaluation results of cross-domain singing voice conversion}

\begin{tabular}{ l c c }
\specialrule{0.1em}{2pt}{2pt}
\multicolumn{1}{c}{\textbf{Method}}               & \textbf{NCC} &  \textbf{COS-SIM} \\ \specialrule{0.1em}{2pt}{2pt}
Baseline1 (cross-domain)          &   $0.926$ &  $0.64$   \\ \hline
Baseline2 (cross-domain)           &   $0.902$  &  $0.57$        \\ \hline
Prop\_hub\_cpc (cross-domain)   & $0.983$ &   $0.67$     \\ \hline
Prop\_hub\_cpc (in-domain)   & $0.987$ &   $0.77$     \\ \specialrule{0.1em}{2pt}{2pt}
\end{tabular}
\label{tab:table4}
\end{table}

\subsubsection{Comparison among different acoustic reference}
\label{ssec:subhead}
In this part, we compare three acoustic reference features. Systems using HuBERT feature, middle hidden feature  PPG-Mid, and mel spectrograms are called  Prop\_hub, Prop\_ppg-mid, Prop\_mel, respectively.

The naturalness, similarity MOS results, and NCC are shown in Table.1. The Prop\_hub model all achieves the best results in terms of naturalness, similarity, and NCC scores. Although Prop\_mel model achieves higher naturalness MOS scores than Prop\_ppg-mid, its similarity and NCC score is lower. This is because the mel spectrograms contains too much singer-dependent information to remove. Now we can conclude that the HuBERT feature is the best choice as the acoustic reference. So in the following experiment, we all use the HuBERT feature as the reference encoder input.

\subsubsection{Ablation study for the CPC module}
\label{ssec:subhead}
To show the effectiveness of the CPC module, we compare the Prop\_hub model with the Prop\_hub\_cpc model, which incorporates the CPC module. As shown by the results in Table.2, incorporating the CPC module in Prop\_hub improves both singing naturalness and similarity. It reveals that CPC module, which serves as the regularization loss applied to the encoder output, can discard local noise and errors and enhance the voice smoothness.

\subsubsection{Comparison with the baseline model}
\label{ssec:subhead}
In this part, we compare our best configuration Prop\_hub\_cpc with the following two state-of-the-art baseline models. Baseline1 is proposed in \cite{ref13}. It consists of a PPG encoder, pitch encoder, speaker encoder, and decoder. Similar to our proposed model, baseline1 is also an end-to-end model, which maps PPGs to waveform directly. Baseline2 is proposed in \cite{ref14}. Apart from modules in Baseline1, Baseline2 incorporates a mel reference encoder. Different from our proposed model, Baseline2 is not an end-to-end model. It maps PPGs to mel spectrograms and then we use a pretrained HiFi-GAN vocoder to generate the waveform.

From the evaluation results shown  in Table.3, we can see our proposed model significantly improves the naturalness and similarity compared with the two baselines, which reveals the the superiority of our proposed method.

\subsubsection{Cross-domain singing voice conversion}
Our work focuses on in-domain singing voice conversion. However, with the VCTK speech dataset involved in training, our model can also make the speaker in the VCTK dataset sing, which is called cross-domain singing voice conversion.  We compare our proposed model with the Baseline1 and Baseline2 model.  We just evaluate the performance via objective metrics normalized cross-correlation (NCC) and Cosine similarity (COS-SIM) between the d-vectors of the converted and the real target audio. The d-vectors are extracted from the pretrained speaker verification model\footnote{\href{https://github.com/resemble-ai/Resemblyzer}{https://github.com/resemble-ai/Resemblyzer}}. Higher the COS-SIM score is,  more similar the timbres are. The evaluation results are presented in Tab.4. Our proposed method still achieves significantly better results than the two baseline models, which is consistent with in-domain singing voice conversion. Meanwhile, the results of cross domain singing voice conversion are inferior to these of in domain singing voice conversion since in domain task is easier.
\section{Conclusion and future work}
\label{sec:Conclusion}
 In this paper, we proposed an end-to-end high-quality singing voice conversion system. It employs a reference encoder to implicitly model the prosody and rhythm of the singing voices. We compare three different features: mel spectrograms, PPG-Mid, and HuBERT feature as the acoustic reference. Experiments show that the HuBERT feature is the best choice. Meanwhile, we incorporate a CPC module to further enhance the smoothness of singing voices. Comprehensive experiments have been performed to verify the superiority of our proposed model. As for future work, we will further improve the performance of cross-domain singing voice conversion.


\newpage
\begin{thebibliography}{99}  

\bibitem{ref1} Kazuhiro Kobayashi, Tomoki Toda, Graham Neubig, Sakriani Sakti, and Satoshi Nakamura, “Statistical singing voice conversion based on direct waveform modification with global variance,” in INTERSPEECH, 2015, pp. 2754–2758.
\bibitem{ref2}  Kazuhiro Kobayashi, Tomoki Toda, Graham Neubig, Sakriani Sakti, and Satoshi Nakamura, “Statistical Singing Voice Conversion with direct Waveform modification based on the Spectrum Differential,” in Interspeech, 2014, pp. 2514-2518
\bibitem{ref3} T. Toda, Y. Ohtani and K. Shikano, "One-to-Many and Many-to-One Voice Conversion Based on Eigenvoices," in ICASSP, 2007, pp. 1249-1252.
\bibitem{ref4} Ganin, Yaroslav, et al, "Domain-Adversarial Training of Neural Networks," The journal of machine learning research, vol.17, pp. 2096–2030, 2016.
\bibitem{ref5}  E. Nachmani and L. Wolf, “Unsupervised singing voice conversion,” in INTERSPEECH, 2019, pp. 2583–2587.
\bibitem{ref6}  C. Deng, C. Yu, H. Lu, C. Weng, and D. Yu, “Pitchnet: Unsupervised singing voice conversion with pitch adversarial network,” in ICASSP, 2020, pp. 7749–7753.
\bibitem{ref7} Y. Luo, C. Hsu, K. Agres, and D. Herremans, “Singing voice conversion with disentangled representations of singer and vocal technique using variational autoencoders,” in ICASSP, 2020, pp. 3277–3281.
\bibitem{ref8} Lu, J., Zhou, K., Sisman, B., Li, H, “Vaw-gan for singing voice conversion with non-parallel training data,” in Asia-Pacific Signal and Information Processing Association Annual Summit and Conference, 2020, pp. 514-519.
\bibitem{ref9} Takahashi, Naoya, Mayank Kumar Singh, and Yuki Mitsufuji. "Hierarchical disentangled representation learning for singing voice conversion." arXiv preprint arXiv:2101.06842, 2021.
\bibitem{ref10} X. Chen, W. Chu, J. Guo, and N. Xu, “Singing voice conversion with non-parallel data,” in MIPR. 2019, pp.292–296.
\bibitem{ref11} Adam Polyak, Lior Wolf, Yossi Adi, and Yaniv Taigman, “Unsupervised cross-domain singing voice conversion,” in INTERSPEECH, 2020, pp. 801-805.
\bibitem{ref12} Songxiang Liu, Yuewen Cao, Na Hu, Dan Su, and Helen Meng, “Fastsvc: Fast cross-domain singing voice conversion with feature-wise linear modulation,” in ICME, 2021, pp. 1-6.
\bibitem{ref13} Guo, Haohan, et al. "Phonetic Posteriorgrams based Many-to-Many Singing Voice Conversion via Adversarial Training." arXiv preprint arXiv:2012.01837, 2020.
\bibitem{ref14} Zhonghao Li, Benlai Tang, Xiang Yin, Yuan Wan, Ling Xu, Chen Shen, and Zejun Ma, “Ppg-based singing voice conversion with adversarial representation learning,” in ICASSP, 2021, pp. 7073–7077.
\bibitem{ref15} Hsu W N, Bolte B, Tsai Y H H, et al, “HuBERT: Self-Supervised Speech Representation Learning by Masked Prediction of Hidden Units[J],” arXiv preprint arXiv:2106.07447, 2021.
\bibitem{ref16} S. Schneider, A. Baevski, R. Collobert, and M. Auli, “wav2vec: Unsupervised pre-training for speech recognition,” arXiv preprint arXiv:1904.05862, 2019.
\bibitem{ref17} A. Baevski, H. Zhou, A. Mohamed, and M. Auli, “wav2vec 2.0: A framework for self-supervised learning of speech representations,” arXiv preprint arXiv:2006.11477, 2020.
\bibitem{ref18} A. Baevski, S. Schneider, and M. Auli, “vq-wav2vec: Self-supervised learning of discrete speech representations,” arXiv preprint arXiv:1910.05453, 2019.
\bibitem{ref19} Wang, Disong, et al. "VQMIVC: Vector Quantization and Mutual Information-Based Unsupervised Speech Representation Disentanglement for One-shot Voice Conversion," in INTERSPEECH, 2021, pp. 1344-1348.
\bibitem{ref20}  Jungil Kong, Jaehyeon Kim, and Jaekyoung Bae, “Hifi-gan: Generative adversarial networks for efficient and high fidelity speech synthesis,” in Advances in Neural Information Processing Systems, 2020, pp. 17022–17033.
\bibitem{ref21}  S. Zhang, M. Lei, Z. Yan, and L. Dai, “Deep-fsmn
for large vocabulary continuous speech recognition,” in
ICASSP, 2018, pp. 5869–5873.
\bibitem{ref22} Wang, Yuxuan, et al. "Tacotron: Towards end-to-end speech synthesis," in INTERSPEECH, 2017, pp. 4006-4010.
\bibitem{ref23}  J. Yamagishi et al., “Cstr vctk corpus: English multi-speaker corpus for cstr voice cloning toolkit,” 2019.
\bibitem{ref24} Z. Duan, H. Fang, B. Li, K. C. Sim, and Y. Wang, “The nus sung and spoken lyrics corpus: A quantitative comparison of singing and speech,” in APSIPA ASC, 2013, pp. 1–9.
\bibitem{ref25}  D. P. Kingma and J. Ba, “Adam: A method for stochastic optimization,” in ICLR, 2015.

\end{thebibliography}
\end{document}